\documentclass[proceedings]{JHEP3} 

\usepackage{epsfig,multicol}			

\newbox\mybox
\newcommand\fverb{\setbox\mybox=\hbox\bgroup\verb}
\newcommand\fverbdo{\egroup\medskip\noindent\fbox{\unhbox\mybox}\ }
\newcommand\fverbit{\egroup\item[\fbox{\unhbox\mybox}]}

\usepackage{graphicx}

\DeclareMathAlphabet{\mathsc}{OT1}{cmr}{m}{sc}

\newcommand{\Sol}  {\mathsc{sol}}
\newcommand{\Atm}  {\mathsc{atm}}
\newcommand{\Sbl}  {\mathsc{sbl}}
\newcommand{\Nev}  {\mathsc{nev}}
\newcommand{\Lsnd} {\mathsc{lsnd}}
\newcommand{\dms}{\Delta m^2_\Sol}
\newcommand{\dma}{\Delta m^2_\Atm}
\newcommand{\dml}{\Delta m^2_\Lsnd}

\conference{International Workshop on Astroparticle and High Energy Physics}

\title{Status of neutrino oscillations II:\\ 
How to reconcile the LSND result?}

\preprint{TUM-HEP-534/03}

\author{\speaker{Thomas Schwetz}\\
  Institut f{\"u}r Theoretische Physik, Physik Department\\
  Technische Universit{\"a}t M{\"u}nchen, 
  James-Franck-Str., D--85748 Garching, Germany
  E-mail: \email{schwetz@ph.tum.de}}

\abstract{%
We present an analysis of the global neutrino oscillation data including the
LSND result in terms of four-neutrino mass schemes and a CPT violating
three-neutrino framework. We find that the strong preference of oscillations
into active neutrinos implied by recent solar+KamLAND as well as atmospheric
neutrino data allows to rule out (2+2) mass schemes at 5.8$\sigma$, whereas
(3+1) schemes are disfavoured at the 3.2$\sigma$ level by short-baseline
experiments. The CPT violating scenario is disfavoured at 3.4$\sigma$ by
KamLAND and atmospheric anti-neutrino data. We conclude that currently
no convincing explanation for the LSND signal exists, and it remains a
puzzle how to reconcile it with the other evidences for neutrino
oscillations.
}

\begin{document}

\section{Introduction}

The large amount of data from the solar~\cite{solar,snosalt},
atmospheric~\cite{skatm,macro}, KamLAND reactor~\cite{kamland} and K2K
accelerator~\cite{k2k} neutrino experiments is beautifully described in terms
of three-flavour neutrino oscillations~\cite{3nu}. To reconcile also the
evidence for oscillations from the LSND experiment~\cite{lsnd} with all the
other data provides a big challenge for neutrino phenomenology. To obtain the
required three neutrino mass-squared differences of different orders of
magnitude it has been proposed to introduce a light sterile
neutrino~\cite{sterile}, or to allow for violation of CPT invariance by
neutrino masses and mixing~\cite{cpt_orig,Barenboim:2002ah}. Here we present
an analysis of the global neutrino oscillation data in terms of four-neutrino
and CPT violating schemes, including data from solar, atmospheric, KamLAND and
K2K neutrino experiments, the LSND experiment, as well as data from
short-baseline (SBL) experiments~\cite{KARMEN,CDHS,bugey} and reactor
experiments~\cite{lbl} reporting no evidence for oscillations. 

After fixing some notations in Sec.~\ref{sec:notation} we show in
Secs.~\ref{sec:2+2}, ¸\ref{sec:3+1}, \ref{sec:global} that for all
possible types of four-neutrino schemes different sub-sets of the data
are in serious disagreement and hence, four-neutrino oscillations do
not provide a satisfactory description of the global oscillation data
including LSND. The details of our calculations can be found in
Refs.~\cite{4nu01,4nu02,solat02}.  In Sec.~\ref{sec:cpt} we show that
also the CPT violating scenario is strongly disfavoured~\cite{cpt}. In
Sec.~\ref{sec:alternatives} we briefly comment on some proposed
alternative solutions for the LSND puzzle, and we summarize our
results in Sec.~\ref{sec:summary}.

\section{Four-neutrino schemes: notations and approximations}
\label{sec:notation}

Four-neutrino mass schemes are usually divided into
the two classes (3+1) and (2+2), as illustrated in Fig.~\ref{fig:4spectra}. 
We note that (3+1) mass spectra include the three-active
neutrino scenario as limiting case. In this case solar and atmospheric
neutrino oscillations are explained by active neutrino oscillations,
with mass-squared differences $\dms$ and $\dma$, and the fourth
neutrino state gets completely decoupled. We will refer to such
limiting scenario as (3+0). In contrast, the (2+2) spectrum is
intrinsically different, as there must be a significant contribution
of the sterile neutrino either in solar or in atmospheric neutrino
oscillations or in both.

\FIGURE[t]{
   \includegraphics[width=0.8\linewidth]{schemes.eps}
      \caption{The six four-neutrino mass spectra, divided into the
        classes (3+1) and (2+2).}
      \label{fig:4spectra}}

Neglecting CP violation, in general neutrino oscillations in
four-neutrino schemes are described by 9 parameters: 3 mass-squared
differences and 6 mixing angles in the unitary lepton mixing matrix.
Here we use a parameterisation introduced in Ref.~\cite{4nu01}, which
is based on physically relevant quantities: the 6 parameters $\dms$,
$\theta_\Sol$, $\dma$, $\theta_\Atm$, $\dml$, $\theta_\Lsnd$ are
similar to the two-neutrino mass-squared differences and mixing angles
and are directly related to the oscillations in solar, atmospheric and
the LSND experiments. For the remaining 3 parameters we use
$\eta_s,\eta_e$ and $d_\mu$. Here, $\eta_s \,(\eta_e)$ is the fraction
of $\nu_s \,(\nu_e)$ participating in solar oscillations, and
($1-d_\mu$) is the fraction of $\nu_\mu$ participating in oscillations
with $\dma$ (for exact definitions see Ref.~\cite{4nu01}). For the
analysis we adopt the following approximations:
\begin{enumerate}
\item
We make use of the hierarchy $\dms \ll \dma \ll \dml$.
This means that for each data set we consider only one mass-squared
difference, the other two are set either to zero or to infinity.
\item
In the analyses of solar and atmospheric data (but not for SBL data) we
set $\eta_e = 1$, which is justified because of strong constraints
from reactor experiments~\cite{bugey,lbl}.
\end{enumerate}

\FIGURE[t]{
 \centering
   \includegraphics[width=0.8\linewidth]{diagram3.eps}
      \caption{%
         Parameter dependence of the different data sets in our
         parameterisation. }
      \label{fig:diagram}
}

Due to these approximations the parameter structure of the four-neutrino
analysis gets rather simple. The parameter dependence of the four data sets
solar, atmospheric, LSND and NEV is illustrated in Fig.~\ref{fig:diagram}. 
The NEV data set contains the experiments KARMEN~\cite{KARMEN},
CDHS~\cite{CDHS}, Bugey~\cite{bugey} and CHOOZ/Palo Verde~\cite{lbl}, reporting
no evidence for oscillations. We see that only $\eta_s$ links solar and
atmospheric data and $d_\mu$ links atmospheric and NEV data. LSND and NEV data
are coupled by $\dml$ and $\theta_\Lsnd$.

\section{(2+2): ruled out by solar and atmospheric data}
\label{sec:2+2}
The strong preference of oscillations into active neutrinos in solar and
atmospheric oscillations leads to a direct conflict in (2+2) oscillation
schemes. We will now show that thanks to recent solar neutrino data
(in particular from SNO~\cite{snosalt}) in combination with the KamLAND
experiment~\cite{kamland}, and the latest SK data on atmospheric
neutrinos~\cite{skatm} the tension in the data has become so strong that (2+2)
oscillation schemes are essentially ruled out.\footnote{Details of our
analyses of the solar, KamLAND and atmospheric neutrino data can be found in
Refs.~\cite{solat02,KL}. For an earlier four-neutrino analysis of solar and
atmospheric data see Ref.~\cite{concha4nu}.} 

In the left panel of Fig.~\ref{fig:etas} we show the $\Delta \chi^2$
from solar neutrino data as a function of $\eta_s$, the parameter
describing the fraction of the sterile neutrino participating in solar
neutrino oscillations.  One can see that the improved determination of
the neutral current event rate from the solar $^8$B flux implied by
the salt enhanced measurement in SNO~\cite{snosalt} leads to a significant
tightening of the constraint on $\eta_s$: the 99\% CL bound improves
from $\eta_s \le 0.44$ for pre-SNO salt to $\eta_s \le 0.31$ including
also the most recent SNO data. Moreover, large values of $\eta_s
\gtrsim 0.5$ are now ruled out by many standard deviations.
The outstanding results of the KamLAND reactor experiment~\cite{kamland}
confirmed the LMA solution of the solar neutrino problem~\cite{KL,sol+KL}.
Apart from this very important result the impact of KamLAND for an admixture
of a sterile neutrino is rather limited (see Fig.~\ref{fig:etas}). Since
KamLAND on its own has no sensitivity to $\eta_s$ the bound is only indirectly
affected due to the better determination of $\dms$ and $\theta_\Sol$. The
combined analysis leads to the 99\% CL bound $\eta_s \le 0.27$.

\FIGURE[t]{
  \includegraphics[width=0.45\linewidth]{etas_sol+kaml.eps}
  \includegraphics[width=0.52\linewidth]{etas_sol+atm.eps}
  \caption{ Left: $\Delta\chi^2$ from solar and KamLAND data as a function of
    $\eta_s$. Right: $\Delta\chi^2_\Sol$, $\Delta\chi^2_{\Atm+\Sbl}$
    and $\bar\chi^2_\mathrm{global}$ as a function of $\eta_s$ in
    (2+2) oscillation schemes.}
  \label{fig:etas}
}

In contrast, in (2+2) schemes atmospheric data imply $\eta_s \ge 0.65$
at 99\% CL, in clear disagreement with the bound from solar data.  In
the right panel of Fig.~\ref{fig:etas} we show the $\Delta\chi^2$ for solar
data and for atmospheric combined with SBL data as a function of
$\eta_s$.  Furthermore, we show the $\chi^2$ of the global data
defined by
\begin{equation}\label{chi2solatm}
\bar\chi^2(\eta_s) \equiv 
\Delta\chi^2_\Sol(\eta_s) + 
\Delta\chi^2_{\Atm + \Sbl}(\eta_s) \,.
\end{equation}
From the figure we find that only at the 4.7$\sigma$ level a value of $\eta_s$
exists, which is contained in the allowed regions of both data sets. This
follows from the $\chi^2$-value $\chi^2_\mathrm{PC} = 22.4$ shown in the
figure. In Refs.~\cite{4nu02,PG} we have proposed a statistical method to
evaluate the disagreement of different data sets in global analyses. The
\textit{parameter goodness of fit} (PG) makes use of the $\bar\chi^2$ defined
in Eq.~(\ref{chi2solatm}). This criterion evaluates the GOF of the
\textit{combination} of data sets, without being diluted by a large number of
data points, as it happens for the usual GOF criterion (for details see
Ref.~\cite{PG}). We find $\chi^2_\mathrm{PG} \equiv \bar\chi^2_\mathrm{min} =
33.2$, which corresponds to 5.7$\sigma$. We conclude that (2+2) mass schemes
are ruled out by the disagreement between the latest solar and atmospheric
neutrino data. This is a very robust result, independent of whether LSND is
confirmed or disproved.\footnote{Sub-leading effects beyond the approximations
adopted here should not affect this result significantly. Allowing for
additional parameters to vary at the percent level might change the {\it
ratio} of some observables~\cite{Pas:2002ah}, however, we expect that the
absolute number of events relevant for the fit will not change substantially.}

\section{(3+1): strongly disfavoured by SBL data}
\label{sec:3+1}

\FIGURE[t]{
  \includegraphics[width=0.6\linewidth]{3p1.eps}
  \caption{Upper bound on $\sin^22\theta_\Lsnd$ from NEV and
  atmospheric neutrino data in (3+1) schemes~\protect\cite{cornering}
  compared to the allowed region from global LSND
  data~\protect\cite{lsnd} and decay-at-rest (DAR) LSND
  data~\protect\cite{Church:2002tc}.}
  \label{fig:3+1}
}

It is known for a long time~\cite{3+1early} that (3+1) mass schemes are
disfavoured by the comparison of SBL disappearance
data~\cite{CDHS,bugey} with the LSND result. In
Ref.~\cite{cornering} we have calculated an upper bound on the LSND
oscillation amplitude $\sin^22\theta_\Lsnd$ resulting from NEV and
atmospheric neutrino data. From Fig.~\ref{fig:3+1} we see that this
bound is incompatible with the signal observed in LSND at the 95\%
CL. Only marginal overlap regions exist between the bound and global
LSND data if both are taken at 99\% CL. An analysis in terms of the
parameter goodness of fit~\cite{4nu02} shows that for most values of
$\dml$ NEV and atmospheric data are compatible with LSND only at more
than $3\sigma$, with one exception around $\dml \sim 6$ eV$^2$, where
the PG reaches 1\%. These results show that (3+1) schemes are strongly
disfavoured by SBL disappearance data.

\section{Comparing (3+1), (2+2) and (3+0) hypotheses}
\label{sec:global}

With the methods developed in Ref.~\cite{4nu01} we are able to
perform a global fit to the oscillation data in the four-neutrino
framework. This approach allows to statistically compare the different
hypotheses. Let us first evaluate the GOF of (3+1) and (2+2) spectra
with the help of the PG method described in Ref.~\cite{PG}. We
divide the global oscillation data in the 4 data sets SOL, ATM, LSND
and NEV. Following Ref.~\cite{4nu02}
we consider
\begin{equation}\label{chi2bar}
\begin{array}{ccl}
\bar\chi^2 &=&
\Delta\chi^2_\Sol(\theta_\Sol,\dms,\eta_s) 
+ \Delta\chi^2_\Atm(\theta_\Atm,\dma,\eta_s,d_\mu) \\
&+& \Delta\chi^2_\Nev(\theta_\Lsnd,\dml,d_\mu,\eta_e) 
+ \Delta\chi^2_\Lsnd(\theta_\Lsnd,\dml) \,,  
\end{array}
\end{equation}
where $\Delta\chi^2_X = \chi^2_X - (\chi^2_X)_\mathrm{min}$ ($X$ =
SOL, ATM, NEV, LSND). In Tab.~\ref{tab:pg} we show the contributions of
the 4 data sets to $\chi^2_\mathrm{PG} \equiv \bar\chi^2_\mathrm{min}$
for (3+1) and (2+2) oscillation schemes. As expected we observe that
in (3+1) schemes the main contribution comes from SBL data due to the
tension between LSND and NEV data in these schemes. For (2+2)
oscillation schemes a large part of $\chi^2_\mathrm{PG}$ comes from
solar and atmospheric data, however, also SBL data contributes
significantly.  This comes mainly from the tension between LSND and
KARMEN~\cite{Church:2002tc}, which does not depend on the mass scheme
and, hence, also contributes in the case of (2+2). Therefore, the
values of $\chi^2_\mathrm{PG}$ in Tab.~\ref{tab:pg} for (2+2) schemes
are higher than the one given in Sec.~\ref{sec:2+2}, where the tension
in SBL data is not included.

\TABLE[t]{
    \begin{tabular}{|c|cccc|c|c|}
    \hline
    & SOL & ATM & LSND & NEV &   $\chi^2_\mathrm{PG}$ & PG \\
    \hline 
(3+1) & 0.0 & 0.4 & 7.2 & 7.0 & 14.6 & $5.6 \times 10^{-3} \: (3.2\sigma)$ \\
(2+2) & 4.1 & 29.2 & 1.2 & 9.7 & 44.1 & $6.1 \times 10^{-9} \: (5.8\sigma)$ \\
    \hline
    \end{tabular}
    \caption{Parameter GOF and the contributions of different data sets to 
    $\chi^2_\mathrm{PG}$ in (3+1) and (2+2) neutrino mass schemes.}
    \label{tab:pg}
}

The parameter goodness of fit is now obtained by evaluating
$\chi^2_\mathrm{PG}$ for 4 DOF~\cite{PG}. This number of degrees of
freedom corresponds to the 4 parameters $\eta_s, d_\mu, \theta_\Lsnd,
\dml$ describing the coupling of the different data sets (see
Eq.~(\ref{chi2bar}) and Fig.~\ref{fig:diagram}). The best GOF is
obtained in the (3+1) case. However, even in this best case the PG is
only 0.56\%. The PG of $6.1\times 10^{-9}$ for (2+2) schemes shows
that these mass schemes are essentially ruled out by the disagreement
between the individual data sets.

Although we have seen that none of the four-neutrino mass schemes can
provide a reasonable good fit to the global oscillation data including
LSND, it might be interesting to consider the \textit{relative} status
of the three hypotheses (3+1), (2+2) and the three-active neutrino
scenario (3+0). This can be done by comparing the $\chi^2$ values of
the best fit point -- which occurs for the (3+1) scheme -- with the
one corresponding to (2+2) and (3+0).  First we observe that (2+2)
schemes are strongly disfavoured with respect to (3+1) with a $\Delta
\chi^2 = 29.5$. For 4 DOF this is equivalent to an exculsion at
4.5$\sigma$. Furthermore, we find that (3+0) is disfavoured with a
$\Delta \chi^2 = 20.0$ (99.95\% CL for 4 DOF) with respect to (3+1).
This reflects the high statistical significance of the LSND result,
since in a (3+0) no effect is predicted for LSND.

\section{CPT violation}
\label{sec:cpt}

The CPT violating solution for the LSND results proposed in
Refs.~\cite{cpt_orig} is based on the observation that the signal in
favour of oscillations in LSND is dominated by the anti-neutrino
signal from decay-at-rest data. The neutrino data from the
decay-in-flight event sample gives a $\nu_\mu\rightarrow \nu_e$
transition probability of $(0.10 \pm 0.16 \pm 0.04)\%$~\cite{lsnd},
which, although consistent with the anti-neutrino signal of
$P_{\bar\nu_\mu \to \bar\nu_e} = (0.264 \pm 0.067 \pm 0.045)\%$, is
also perfectly consistent with the absence of neutrino oscillations in
LSND. Therefore, it has been proposed to adopt different masses and
mixings for neutrinos and anti-neutrinos to reconcile the LSND
result~\cite{cpt_orig,strumia} without introducing a sterile
neutrino. In such a scheme solar and atmospheric neutrino oscillations
are driven by $\Delta m^2_\Sol$ and $\Delta m^2_\Atm$, and atmospheric
and LSND anti-neutrino oscillations by $\Delta\overline{m}^2_\Atm$ and
$\Delta\overline{m}^2_\Lsnd$.

\FIGURE[t]{
  \includegraphics[width=0.6\linewidth]{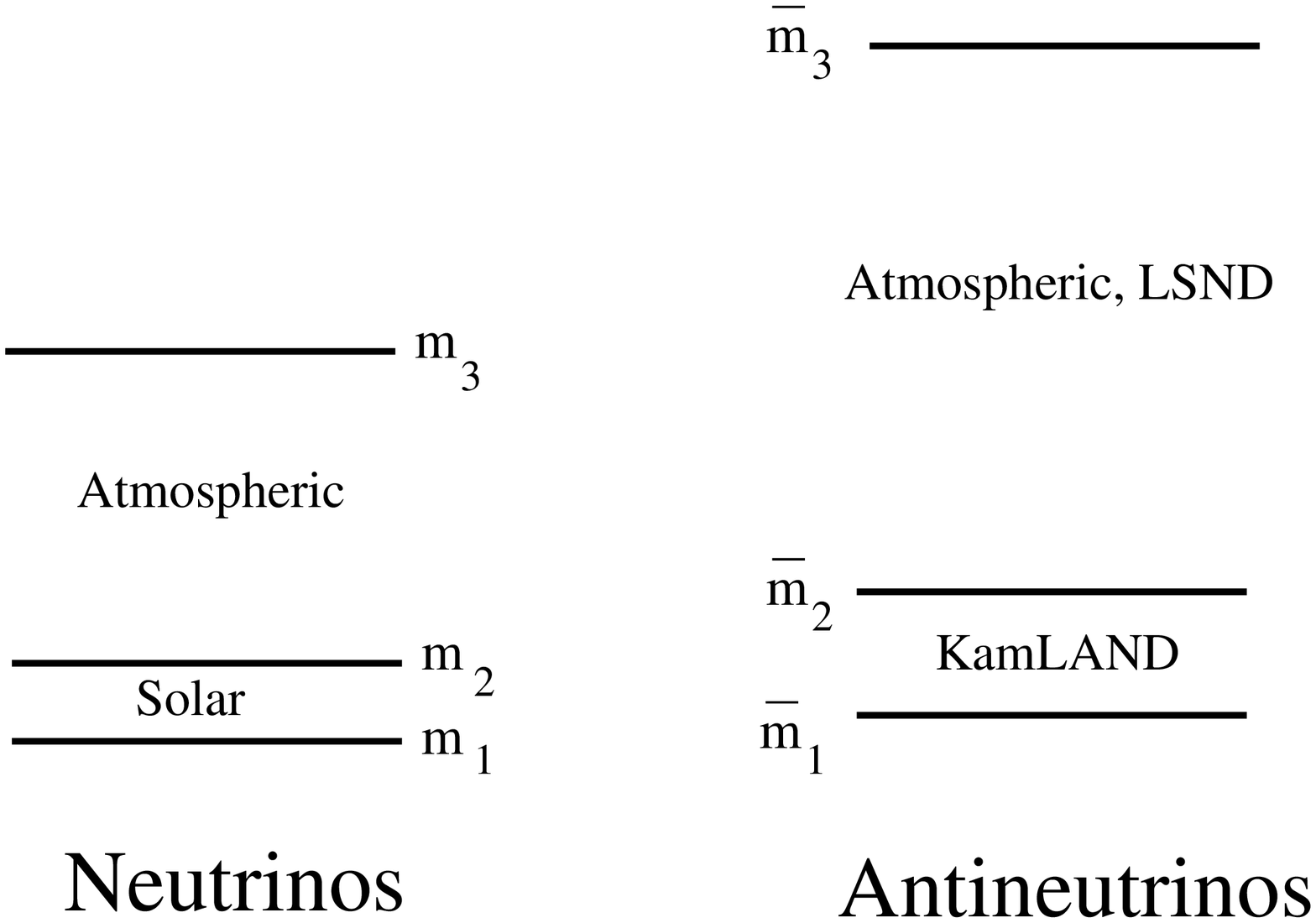}
  \caption{CPT violating neutrino mass scheme proposed in
  Ref.~\protect\cite{Barenboim:2002ah}.} 
  \label{fig:cpt_scheme}
}

This scenario is strongly affected by the KamLAND results on
anti-neutrino disappearance, which in combination with the constraints
from CHOOZ indicate oscillations with $\Delta\overline{m}^2 \lesssim
10^{-4}$~eV$^2$. Since this requires a third mass scale in the
anti-neutrino sector the CPT violating solution seems to be
disfavoured. However, in Ref.~\cite{Barenboim:2002ah} it was claimed
that the mass spectrum shown in Fig.~\ref{fig:cpt_scheme} still may
provide a fit to the data, although no anti-neutrino mass-squared
difference in the range required for atmospheric oscillations is
present. This claim is based on the fact that Super-Kamiokande cannot
distinguish between neutrinos and anti-neutrinos, and the atmospheric
data sample is dominated by neutrinos by a factor 2--4 over
anti-neutrinos. 

In Ref.~\cite{cpt} we have performed an anlysis of the global
oscillation data in the framework of the CPT violating neutrino
scheme. First we have analized all data except LSND -- from
solar, atmospheric, reactor (KamLAND, CHOOZ, Bugey) and K2K
experiments -- in a general three-flavour context, by allowing for
different neutrino and anti-neutrino oscillation parameters. 
We found that there is no evidence for any CPT violation in that data
set. In fact, the best fit point occurs very close to CPT
conservation, and the best fit in the CPT conserving case has 
$\Delta \chi^2$ of only 0.5.

In Fig.~\ref{fig:cpt} we show the allowed regions of the all-but-LSND
data compared to the LSND allowed regions, projected onto the plane of
the LSND oscillation parameters $\Delta m^2_\Lsnd = \Delta
\overline{m}^2_{31}$ and $\sin^2 2\theta_\Lsnd =
\sin^2\overline{\theta}_{23} \sin^2 2\overline{\theta}_{13}$.  This
figure illustrates that below 3$\sigma$ CL there is no overlap between
the allowed region of the LSND analysis and the all-but-LSND one, and
that for this last one the region is restricted to $\Delta m^2_\Lsnd <
0.02$~eV$^2$.  At higher CL values of $\Delta m^2_\Lsnd \sim {\cal
O}(\mathrm{eV}^2)$ become allowed -- as determined mainly by the
constraints from Bugey -- and an agreement becomes possible. We find
that in the neighbourhood of $\Delta m^2_\Lsnd = 0.9$~eV$^2$
and $\sin^2 2\theta_\Lsnd = 0.01$ the LSND and the all-but-LSND
allowed regions start having some marginal agreement slightly above
3$\sigma$ CL (at $\Delta\chi^2 = 12.2$).  A less fine-tuned agreement
appears at 3.3$\sigma$ CL ($\Delta\chi^2\sim 14)$ for 
$\Delta m_\Lsnd \gtrsim 0.5$~eV$^2$ and $\sin^2 2\theta_\Lsnd\lesssim 0.01$.

\FIGURE[t]{
  \includegraphics[width=0.6\linewidth]{cpt}
  \caption{90\%, 95\%, 99\%, and $3\sigma$ CL allowed regions
      (filled) in the ($\Delta m^2_\Lsnd$, $\sin^2 2\theta_\Lsnd$)
      plane required to explain the LSND signal together with the
      corresponding allowed regions from our global analysis of
      all-but-LSND data. The contour lines correspond to $\Delta\chi^2
      = 13$ and 16 (3.2$\sigma$ and 3.6$\sigma$, respectively).}
  \label{fig:cpt}
}

Alternatively the quality of the joint description of LSND and all the
other data can be evaluated by performing a global fit based on the
total $\chi^2$-function $\chi^2_\mathrm{tot} = \chi^2_\mathrm{all-but-LSND}
+ \chi^2_\mathrm{LSND}$, and applying a goodness-of-fit test. The best
fit point of the global analysis is $\sin^22\theta_\Lsnd = 6.3 \times
10^{-3}$ and $\Delta m^2_\Lsnd = 0.89$~eV$^2$.  In the following we
will again use the parameter goodness-of-fit~\cite{PG}, which is
particularly suitable to test the compatibility of independent data
sets.  Applying this method to the present case we consider the
statistic
\begin{equation} 
    \bar\chi^2 = \Delta\chi^2_\mathrm{all-but-LSND}(\mathrm{b.f.}) +
    \Delta\chi^2_\Lsnd(\mathrm{b.f.}) \,,\label{eq:chi2PG}
\end{equation}
where b.f.\ denotes the global best fit point.  The $\bar\chi^2$ of
Eq.~(\ref{eq:chi2PG}) has to be evaluated for 2 dof, corresponding to
the two parameters $\sin^22\theta_\Lsnd$ and $\Delta
\overline{m}^2_{31}= \Delta m^2_\Lsnd$ coupling the two data sets.
From $\Delta\chi^2_\mathrm{all-but-LSND} = 12.7$ and
$\Delta\chi^2_\mathrm{LSND} = 1.7$ we obtain $\bar\chi^2 = 14.4$
leading to the marginal parameter goodness-of-fit of $7.5\times
10^{-4}$, corresponding to 3.4$\sigma$.

These results show that atmospheric data is sensitive enough to
anti-neutrino oscillations, and hence, no reasonable fit can be
obtained without a mass-squared difference of the order 
$10^{-3}$~eV$^2$ for anti-neutrinos. Consequently the CPT violation
scenario is disfavoured by global anti-neutrino data from reactor,
atmospheric and the LSND experiments.

\section{Alternative solutions}
\label{sec:alternatives}

In this section we briefly mention some alternative solutions which
have been proposed to explain the result from LSND. 

\begin{itemize}
\item
In Ref.~\cite{Babu:2002ic} a lepton number violating decay of the muon
has been considered as the source for the LSND signal. In this
model the $\bar\nu_e$ event excess in LSND stems from the decay $\mu^+
\to e^+ + \bar\nu_e + \bar\nu_x$, where $\bar\nu_x$ can be an
anti-neutrino of any flavour. This scenario seems to be in
disagreement with the non-observation of any $\bar\nu_e$ excess in
KARMEN.
\item
Recently an explanation in terms of five neutrinos has been
proposed~\cite{Sorel:2003hf}. In such a (3+2) scheme solar and
atmospheric data are explained dominantly by active oscillations,
similar to the (3+1) case. The fit of SBL data is improved with
respect to (3+1) by involving two large mass splittings $\Delta
m^2_{41} \sim 0.9$ eV$^2$ and $\Delta m^2_{51} \sim 20$
eV$^2$. In this case it seems to be very dificult to reconcile these
large neutrino masses with constraints from cosmology~\cite{cosmology}.
\item
In Ref.~\cite{4nu+cpt} it has been pointed out that involving a
sterile neutrino {\it and} CPT violation simultaneously all the data
can be explained. Once such drastical modifications of standard
physics are accepted, one may adopt a (3+1) like mass scheme, even with
identical masses for neutrinos and anti-neutrinos, but allowing for
different mixings. Since CDHS constrains only SBL muon neutrino
disappearance (not anti-neutrinos), whereas Bugey restricts
$\bar\nu_e$ mixing, a small violation of CPT would suffice to get
around the SBL bounds and explain the LSND signal.
\end{itemize}

\section{Summary}
\label{sec:summary}

Performing a global analysis of current neutrino oscillation data we find that
neither for four-neutrino schemes nor for CPT violation in a three-neutrino
framework a satisfactory fit to the data is obtained.
\begin{itemize}
\item
The strong rejection of non-active oscillation in the solar+KamLAND
and atmospheric neutrino data rules out (2+2) schemes, independent of whether
LSND is confirmed or not. Using an improved goodness of fit method especially
sensitive to the combination of data sets we find that (2+2) schemes
are ruled out at the 5.8$\sigma$ level.
\item 
(3+1) spectra are disfavoured by the disagreement of LSND with
short-baseline disappearance data, leading to a marginal GOF of
$5.6\times 10^{-3}$ (3.2$\sigma$).  If LSND should be confirmed it
would be very desirable to have more data on $\nu_e$ and/or $\nu_\mu$
SBL disappearance to decide about the status of (3+1). In that case a
positive signal is predicted right at the sensitivity edge of existing
experiments. 
\item
Also in the case of the CPT violating three-neutrino mass scheme no
satisfactory fit can be obtained. This scheme is disfavoured by
KamLAND, CHOOZ, Bugey and atmospheric anti-neutrino data at the
3.4$\sigma$ level.
\end{itemize}

Alternative solutions, which have been proposed to reconcile the
LSND result, seem to be in disagreement with some data, or require a
very radical modification of standard physics. We conclude that
currently no convincing explanation for the LSND result exists, and it
remains a puzzle how to reconcile this evidence with the rest of the
global data. Therefore, it is very important to settle this issue
experimentally. A confirmation of the LSND signal by the MiniBooNE
experiment~\cite{miniboone} would be very exciting and would require
some new ideas.

\acknowledgments

I would like to thank M.C.~Gonzalez-Garcia, M.~Maltoni,
M.A.~T{\'o}rtola and J.W.F.~Valle for collaboration on the work
presented here. Furthermore, I thank the organisers for the very
interesting workshop and for financial support. This work was
supported by the ``Sonderforschungsbereich 375-95 f{\"u}r
Astro-Teilchenphysik'' der Deutschen Forschungsgemeinschaft.

\end{document}